\documentclass{webofc}
\usepackage[varg]{txfonts}   % Web of Conferences font
\usepackage[rightcaption,raggedright]{sidecap}
\usepackage{ragged2e}
\usepackage[colorlinks=true,
			linktocpage=true,
			linkcolor=blue,
			citecolor=red,
			urlcolor=blue]{hyperref}
\newcommand{\nn}{\nonumber\\}
\newcommand{\pder}[2]{\frac{\partial#1}{\partial#2}}
\def\tR{\tau_R}
\def\L{{\cal L}}
\def\M{{\cal M}}
\newcommand{\del}{\partial}
\begin{document}
%
% \title{Applicability of higher-order hydrodynamics in heavy-ion collisions}
\title{Why are hydrodynamic theories applicable beyond the hydrodynamic regime?}
%
% subtitle is optional

\author{
\vspace{-.4cm}
\firstname{Sunil} \lastname{Jaiswal}
    \inst{1}\fnsep\thanks{\email{jaiswal.61@osu.edu}} \and
\firstname{Jean-Paul} \lastname{Blaizot}
    \inst{2}\fnsep\thanks{\email{jean-paul.blaizot@ipht.fr}} \and
\firstname{Rajeev S.} \lastname{Bhalerao}
    \inst{3} \and
\firstname{Zenan} \lastname{Chen}
    \inst{4} \and
\firstname{Amaresh} \lastname{Jaiswal}
    \inst{5} \and
\firstname{Li} \lastname{Yan}
    \inst{4}
}
\institute{
Department of Physics, The Ohio State University, Columbus, Ohio 43210-1117, USA 
\and
Institut de Physique Th{\'e}orique, Universit\'e Paris Saclay, CEA, CNRS, F-91191 Gif-sur-Yvette, France
\and
Department of Physics, Indian Institute of Science Education and Research (IISER), Homi Bhabha Road, Pashan, Pune 411008, India
\and
Institute of Modern Physics Fudan University, 220 Handan Road, 200433, Shanghai, China
\and
School of Physical Sciences, National Institute of Science Education and Research, An OCC of Homi Bhabha National Institute, Jatni-752050, India
}

\abstract{%
We present an alternative approach to deriving second-order non-conformal hydrodynamics from the relativistic Boltzmann equation. We demonstrate how constitutive relations for shear and bulk stresses can be transformed into dynamical evolution equations, resulting in Israel-Stewart-like (ISL) hydrodynamics. To understand the far-from-equilibrium applicability of such ISL theories, we investigate the one-dimensional boost-invariant Boltzmann equation using special moments of the distribution function for a system with finite particle mass. Our analysis reveals that the mathematical structure of the ISL equations is akin to that of moment equations, enabling them to approximately replicate even the collisionless dynamics. We conclude that this particular feature is important in extending the applicability of ISL theories beyond the hydrodynamic regime.
}
\maketitle
%

%%%%%%%%%%%%%%%%%%%%%%%%%%%%%%%%%%%%%%
\vspace{-.5cm}
\section{Hydrodynamics and Israel-Stewart theory} 
%%%%%%%%%%%%%%%%%%%%%%%%%%%%%%%%%%%%%%
%
Kinetic theory provides a microscopic description capable of accurately describing a far-from-equilibrium systems as well as the collective dynamics of a system close to equilibrium, and has been employed extensively to formulate the theory of relativistic dissipative hydrodynamics \cite{Florkowski:2017olj}. The relativistic Boltzmann equation for a gas of massive particles in the relaxation-time approximation is given by: $p^\mu \partial_\mu f = -\frac{(u\cdot p)}{\tau_{\rm R}} \left(f - f_{\rm eq}\right)$, where $\tau_{\rm R}$ denotes the relaxation time. By employing Chapman-Enskog-like expansion of $f(x,p)$, we derive the correction to the equilibrium distribution function up to second order in gradients as:
{\small
\begin{align}\label{deltaf2}
    \delta f &= \frac{\tau_{\rm R}}{T(u\cdot p)} \left[ \left[ m^2 -(1-3c_s^2) (u\cdot p)^2 \right] \frac{\theta}{3} + p^\mu p^\nu \sigma_{\mu\nu} \right] f_{\rm eq}
\nn
    &- \tau_{\rm R} D \!\left[\! \frac{\tau_{\rm R}}{T(u\cdot p)} 
        \!\left\{ \left[m^2 \!-\! (1-3c_s^2) (u\cdot p)^2 \right] \frac{\theta}{3} \!+\! p^\mu p^\nu \sigma_{\mu\nu} \!\right\}\! f_{\rm eq} \!\right]
    \!-\! \frac{\tau_{\rm R}}{u\cdot p} p^\mu \nabla_\mu 
    \!\left[\! \frac{\tau_{\rm R}}{T} \!\left\{ (u\cdot p) c_s^2\, \theta \!+\! \frac{p^\alpha p^\nu}{u\cdot p} (\nabla_\nu u_\alpha ) \!\right\}\! f_{\rm eq} \!\right] 
\nn
    &+ \frac{\tau_{\rm R}}{T(\epsilon+P)} \left[ p^\mu (\nabla_\mu \Pi) -\Pi\, p^\mu \dot{u}_\mu + (u\cdot p) c_s^2\, \Pi\, \theta - (u\cdot p) c_s^2\, \pi^{\mu\nu} \sigma_{\mu\nu} - p^\mu \Delta_{\mu\alpha} \partial_\beta \pi^{\alpha\beta} \right] f_{\rm eq} ,
\end{align}
}%
where we have used the standard notations and definitions~\cite{Jaiswal:2014isa}. The out of equilibrium temperature `$T$' and fluid velocity `$u^\mu$' are defined using the Landau matching and frame conditions.

Using the definitions of  bulk and shear stresses,
\begin{equation}\label{VT}
    \Pi \equiv -\frac{1}{3}\Delta_{\alpha\beta} \!\int\! dP \, p^\alpha p^\beta\, \delta f, \qquad 
    \pi^{\mu\nu} \equiv \Delta^{\mu\nu}_{\alpha\beta} \!\int\! dP \, p^\alpha p^\beta\, \delta f \,, 
\end{equation}
the second-order constitutive relations using  Eq.~\eqref{deltaf2} are obtained to be:
\begin{align}
\label{bulkGE2}
\frac{\Pi}{\tau_{\rm R}} =& -\beta_\Pi \theta - D[-\tau_{\rm R} \beta_\Pi \theta] + c_s^2 \Pi \theta - c_s^2 \pi^{\mu\nu} \sigma_{\mu\nu} 
+ \frac{5}{9} \frac{\tau_{\rm R}}{T} \left[ 3 c_s^2 (I_{31}+I_{32}) + (5 I_{32}+7 I_{33}) \right] \theta^2
\nn
& +\frac{4}{3} \frac{\tau_{\rm R}}{T} (2 I_{32}+7 I_{33}) \sigma^{\mu\nu} \sigma_{\mu\nu} \,,
\\ \label{shearGE2}
\frac{\pi^{\mu\nu} }{\tau_{\rm R}} =&\, 2 \beta_\pi \sigma^{\mu\nu} - \Delta_{\alpha\beta}^{\mu\nu} 
D\left[2 \tau_{\rm R} \beta_\pi \sigma^{\mu\nu}\right] 
+ 4\tau_{\rm R} \beta_\pi \sigma_{\rho}^{\langle\mu} \omega^{\nu \rangle\rho} 
- 4 \frac{\tau_{\rm R}}{T}  \left(I_{32}+2I_{33}\right) \sigma_{\rho}^{\langle\mu} \sigma^{\nu \rangle\rho}
\nn
&- \frac{\tau_{\rm R}}{T} \left[2c_s^2 (I_{31}+I_{32}) +\frac{4}{3} (5 I_{32}+7 I_{33})\right] \sigma^{\mu\nu} \theta \,.
\end{align}
Equations~\eqref{bulkGE2} and \eqref{shearGE2} are exact up to second order, and along with energy-momentum conservation equations describe the evolution of the hydrodynamic fields, i.e., temperature and fluid velocity. These equations are the second-order hydrodynamic equations in the traditional sense, where no new fields are introduced. The resulting equations are however acausal.

Phenomenological Israel-Stewart theory avoids the acausality issue by promoting the dissipative stresses to dynamical degrees of freedom \cite{Israel:1976tn, Israel:1979wp}. Israel-Stewart-like (ISL) hydrodynamic equations, commonly refereed to in the heavy-ion community as `second-order hydrodynamics' and used in hydrodynamic simulation of heavy-ion collisions, can be obtained by replacing the gradients $\theta$ and $\sigma_{\mu\nu}$ in Eqs.~\eqref{bulkGE2} and \eqref{shearGE2} using the first order constitutive relations for dissipative stresses, i.e.,  $\theta \to -\frac{\Pi}{\tau_{\rm R} \beta_\Pi}$ and $\sigma^{\mu\nu} \to \frac{\pi^{\mu\nu}}{2 \tau_{\rm R} \beta_\pi}$. The resulting equations are now transformed into the dynamical evolution equations:
\begin{align}
\label{IS_BULK}
    \tau_{\rm R} \dot{\Pi}  +\Pi &=  -\zeta \theta - \tau_{\rm R}\delta_{\Pi \Pi}  \Pi \theta + \tau_{\rm R} \lambda_{\Pi \pi} \pi^{\mu \nu } \sigma_{\mu \nu } ,
\\ \label{IS_SHEAR}
    \tau_{\rm R}\dot{\pi}^{\langle\mu\nu \rangle}+\pi^{\mu\nu}  &= 2 \eta  \sigma^{\mu\nu} +2\tau_{\rm R} \pi_{\alpha}^{\langle \mu} \omega^{\nu \rangle\alpha} 
    -\tau_{\rm R} \tau_{\pi\pi} \pi_{\alpha}^{\langle\mu} \sigma^{\nu \rangle\alpha} 
    -\tau_{\rm R} \frac{\mathcal{S}_\pi}{2\beta_\pi} \pi^{\mu\nu} \theta 
    +\tau_{\rm R} \frac{\mathcal{S}_\Pi}{\beta_\Pi} \Pi \sigma^{\mu\nu}.
\end{align}
(See Ref.~\cite{Jaiswal:2014isa} for expressions of the transport coefficients.) The above equations are exact up to second order and are not a priori expected to work in regimes of large gradients. Surprisingly, comparison of such ISL equations with kinetic theory in flow profiles amenable to semi-analytic treatment have shown that they provide a good approximation of the exact dynamics even in far-off-equilibrium regimes \cite{Jaiswal:2014isa, Denicol:2014mca, Chattopadhyay:2018apf, Chattopadhyay:2021ive, Jaiswal:2021uvv}. We shall further examine this in the next section.

It is important to note that the last two terms in Eq.~\eqref{IS_SHEAR} originate from the last term in Eq.~\eqref{shearGE2}, and therefore must satisfy the constraint,
\begin{equation}\label{cond1}
\mathcal{S}_\pi + \mathcal{S}_\Pi = \frac{1}{T} \left[2c_s^2 (I_{31}+I_{32}) +\frac{4}{3} (5 I_{32}+7 I_{33})\right].
\end{equation}
The choice,
\begin{align}
\mathcal{S}_\pi &= \frac{2}{3T} (5 I_{32}+7 I_{33}), \qquad
\mathcal{S}_\Pi = \frac{2}{T} \left[c_s^2 (I_{31}+I_{32}) +\frac{1}{3} (5 I_{32}+7 I_{33})\right],
\end{align}
respects condition~\eqref{cond1} and provides the expressions for $\delta_{\pi\pi}=\frac{\mathcal{S}_\pi}{2\beta_\pi}$ and $\lambda_{\pi\Pi}=\frac{\mathcal{S}_\Pi}{\beta_\Pi}$ obtained in prior works \cite{Jaiswal:2014isa}, making Eqs.~\eqref{bulkGE2} and \eqref{shearGE2} identical to those obtained in \cite{Jaiswal:2014isa}. We emphasize that there is no unique method to determine the coefficients $\mathcal{S}_\pi$ and $\mathcal{S}_\Pi$. Any combination satisfying constraint~\eqref{cond1} is a valid choice. Nevertheless, various combinations impact the shear-bulk coupling and exert influence in the far-off-equilibrium regime. For instance, the selection of these terms as proposed in Ref.~\cite{Jaiswal:2022udf} is better suited for Bjorken geometry.

%%%%%%%%%%%%%%%%%%%%%%%%%%%%%%%%%%%%%%
\section{Moments of Boltzmann equation} 
%%%%%%%%%%%%%%%%%%%%%%%%%%%%%%%%%%%%%%
%
Simulations of ultra-relativistic collisions based on ISL theories have been successful in describing the final-state observables. However, hydrodynamics is expected to break down for such systems as the evolution of the formed matter at early stages is affected by large spatial and temporal gradients. To better understand this unexpected effectiveness, we consider an idealization of the early stages of a high-energy heavy-ion collision, where the produced matter undergoes a boost-invariant expansion along the $z$ direction \cite{Bjorken:1982qr}. The kinetic equation describing the evolution of the distribution function, in the relaxation-time approximation is \cite{Baym:1984np}
\begin{equation}\label{RTA}
\left(\frac{\partial}{\partial\tau} - \frac{p_z}{\tau} \frac{\partial}{\partial p_z}\right) f(\tau, p) = -\frac{f(\tau, p)- f_{\rm eq}{(p_0}/T)}{\tau_R},
\end{equation}
where $\tau$ is the proper time, and $p_0=\sqrt{m^2+p^2}$ is the energy of a particle with mass $m$ and momentum $p$.

For a system undergoing Bjorken expansion, the components of energy-momentum tensor ($\epsilon, P_T$ and $P_L$) can be defined in terms of the moments \cite{Blaizot:2017ucy, Blaizot:2019scw}
\begin{equation}\label{L_mom}
\L_n \equiv  \int dP\, p_0^2 \ P_{2n}(\cos \psi) \ f(\tau, p) \,,
\end{equation}
where $P_{2n}$ is the Legendre polynomial and $\cos\psi \equiv p_z/p_0$. The first two $\L_n$-moments are $\L_0 =\epsilon$ and $\L_1 = \left(3P_L -\epsilon\right)/2$. The transverse pressure involves in addition the trace of the energy-momentum tensor, $P_T = \frac{1}{3} \left( \L_0 -\L_1 -\frac{3}{2}  T^\mu_\mu \right)$, which cannot be expressed solely in terms of the $\L_n$-moments. This requires another type of moments, which we define as \cite{Jaiswal:2022udf}
\begin{equation} \label{M_mom}
\M_n \equiv m^2 \int dP\, P_{2n}(\cos \psi) \ f(\tau, p).
\end{equation}
The moment $\M_0$ is equal to the trace of the energy-momentum tensor $T^\mu_\mu$. The bulk viscous pressure ($\Pi$), and a single independent shear stress tensor component ($\phi$) can be expressed in terms of the moments and the equilibrium pressure ($P$),
\begin{equation}\label{eq_isoP_phi}
P+\Pi = \frac{1}{3} \left( \L_0 - \M_0 \right) , \qquad \phi = -\frac{2}{3} \left( \L_1 + \frac{\M_0}{2} \right).
\end{equation}
Using the relations among the Legendre polynomials and the definitions~\eqref{L_mom} and \eqref{M_mom} of the moments, the kinetic equation~\eqref{RTA} can be recast into a hierarchy of coupled equations:
\begin{align}
\label{L_eqn}
    \pder{\L_n}{\tau} =& -\frac{1}{\tau} \left( a_n \L_n + b_n \L_{n-1} + c_n \L_{n+1} \right) - \frac{\left( \L_n - \L_n^{\rm eq} \right)}{\tR} \left(1- \delta_{n,0} \right) \, ,
\\ \label{M_eqn}
    \pder{\M_n}{\tau} =& -\frac{1}{\tau} \left(a'_n \M_n + b'_n \M_{n-1} + c'_n \M_{n+1}\right) - \frac{\left( \M_n - \M_n^{\rm eq} \right)}{\tR},
\end{align}
where the coefficients $a_n,b_n,c_n$ and $a'_n,b'_n,c'_n$ are real constants \cite{Jaiswal:2022udf}. The equations for the lowest three moments $\L_0$, $\L_1$, and $\M_0$ fully represent the evolution of energy-momentum tensor. The last terms (proportional to the collision rate $1/\tau_{\rm R}$) in above equations arise from the RTA collision kernel and capture the effect of the collisions. Without these terms, Eqs.~(\ref{L_eqn}, \ref{M_eqn}) describe the free streaming (collisionless) regime, where the moments evolve as power laws. The collision term in the above equations produces a damping of the moments and drives the system towards local equilibrium.

The $\L_n$ moment equations are decoupled from the $\M_n$ moment equations. To proceed, it is instructive to write the $\L_0$ equation in terms of the variable $g_0=\frac{\tau}{\L_0} \frac{\del \L_0}{\del \tau}$,
\begin{equation}\label{eq:betafct}
-\beta(g_0,w) = g_0^2+g_0\left [ a_0+a_1+w \right]+a_0a_1 -c_0b_1 +a_0 w
-c_0 c_1 \frac{\L_2}{\L_0}-\frac{c_0}{2} w \left(1-3\frac{P}{\epsilon}\right),
\end{equation}
where $\beta(g_0,w) = w \frac{{\rm d} g_0}{{\rm d} w}$ and $w \equiv \tau/\tau_{\rm R}$. Note that in writing the above equation, the time derivative of $\L_1$ was crucial and it coupled the above equation to higher moments. In collisionless regime, $w \ll 1$, the function $\beta(g_0,w)$ is dominated by the terms that do not depend on $w$. The zeroes of this function correspond to two fixed points of the free-streaming evolution, which have the exact values of $g_0 = -1, -2$. A naive truncation where one simply ignores the term proportional to $\L_2$ in Eq.~(\ref{eq:betafct}), the fixed point values are respectively $-0.93$ and $-2.21$, approximately capturing the location of these fixed points. The hydrodynamic fixed point, $g_*=-a_0 + \frac{c_0}{2}\left(1-3\frac{P}{\epsilon}\right)$, is reached when the collision rate dominates the expansion, i.e. $w\gg 1$, and its location is exactly captured even by this truncated equations. This implies that even the naively truncated moment equations capture approximately the exact dynamics even in collisionless regime (see \cite{Jaiswal:2022udf, Blaizot:2023yrh} for more details).

The equations for the moments, $\L_1$ and $\M_0$ can be reformulated in terms of equations for shear and bulk pressures using relations~\eqref{eq_isoP_phi}. The ISL equations~(\ref{bulkGE2},\ref{shearGE2}) corresponding to Bjorken flow emerge naturally from the moment equations by truncating the moment hierarchy, and they retain the approximate values of the previously mentioned fixed points \cite{Jaiswal:2022udf}. The fact that Israel-Stewart equations apparently allow ``hydrodynamics'' to work in far-off-equilibrium regimes has little to do with hydrodynamics, but rather with the fact that the structure of Israel-Stewart equations is similar to that of the moment equations. Thus they are able to approximately capture some of the features of the collisionless regime.

% Bibliography:

% BibTeX or Biber users please use (the style is already called in the class, ensure that the "woc.bst" style is in your local directory)
%%%%%%%%%%%%%%%%%%%%%%%%
\vspace*{-3mm}
\bibliography{reference.bib}

\begin{thebibliography}{14}

\bibitem{Florkowski:2017olj}
W.~Florkowski, M.P. Heller, M.~Spalinski, Rept. Prog. Phys. \textbf{81}, 046001
  (2018), \texttt{1707.02282}

\bibitem{Jaiswal:2014isa}
A.~Jaiswal, R.~Ryblewski, M.~Strickland, Phys. Rev. C \textbf{90}, 044908
  (2014), \texttt{1407.7231}

\bibitem{Israel:1976tn}
W.~Israel, Annals Phys. \textbf{100}, 310 (1976)

\bibitem{Israel:1979wp}
W.~Israel, J.~Stewart, Annals Phys. \textbf{118}, 341 (1979)

\bibitem{Denicol:2014mca}
G.S. Denicol, W.~Florkowski, R.~Ryblewski, M.~Strickland, Phys. Rev. C
  \textbf{90}, 044905 (2014), \texttt{1407.4767}

\bibitem{Chattopadhyay:2018apf}
C.~Chattopadhyay, U.~Heinz, S.~Pal, G.~Vujanovic, Phys. Rev. C \textbf{97},
  064909 (2018), \texttt{1801.07755}

\bibitem{Chattopadhyay:2021ive}
C.~Chattopadhyay, S.~Jaiswal, L.~Du, U.~Heinz, S.~Pal, Phys. Lett. B
  \textbf{824}, 136820 (2022), \texttt{2107.05500}

\bibitem{Jaiswal:2021uvv}
S.~Jaiswal, C.~Chattopadhyay, L.~Du, U.~Heinz, S.~Pal, Phys. Rev. C
  \textbf{105}, 024911 (2022), \texttt{2107.10248}

\bibitem{Jaiswal:2022udf}
S.~Jaiswal, J.P. Blaizot, R.S. Bhalerao, Z.~Chen, A.~Jaiswal, L.~Yan, Phys.
  Rev. C \textbf{106}, 044912 (2022), \texttt{2208.02750}

\bibitem{Bjorken:1982qr}
J.D. Bjorken, Phys. Rev. \textbf{D27}, 140 (1983)

\bibitem{Baym:1984np}
G.~Baym, Phys. Lett. B \textbf{138}, 18 (1984)

\bibitem{Blaizot:2017ucy}
J.P. Blaizot, L.~Yan, Phys. Lett. B \textbf{780}, 283 (2018),
  \texttt{1712.03856}

\bibitem{Blaizot:2019scw}
J.P. Blaizot, L.~Yan, Annals Phys. \textbf{412}, 167993 (2020),
  \texttt{1904.08677}

\bibitem{Blaizot:2023yrh}
J.P. Blaizot, \emph{{Emergence of hydrodynamics in expanding relativistic
  plasmas}}, in \emph{{Excited QCD 2022}} (2023), \texttt{2309.14124}

\end{thebibliography}
%%%%%%%%%%%%%%%%%%%%%%%%

\end{document}